\documentclass[prb,aps,twocolumn,showpacs,superscriptaddress,floatfix]{revtex4}
\usepackage{epsfig}
\usepackage{amsmath}
\usepackage{amssymb}
\usepackage{amsfonts}

\begin{document}

\title{Giant mesoscopic fluctuations of the elastic cotunneling
thermopower of a single-electron transistor}
\author{A.~S.~Vasenko}
\affiliation{University Grenoble Alpes, CNRS, LPMMC, F-38000 Grenoble,
France}
\author{D.~M.~Basko}
\affiliation{University Grenoble Alpes, CNRS, LPMMC, F-38000 Grenoble,
France}
\author{F.~W.~J.~Hekking}
\affiliation{University Grenoble Alpes, CNRS, LPMMC, F-38000 Grenoble,
France} \affiliation{Institut Universitaire de France, 103, bd
Saint-Michel 75005 Paris, France}

\begin{abstract}
We study the thermoelectric transport of a small metallic
island weakly coupled to two electrodes by tunnel junctions.
In the Coulomb blockade regime, in the case when the ground
state of the system corresponds to an even number of electrons
on the island, the main mechanism of electron transport at the
lowest temperatures is elastic cotunneling.
In this regime, the transport coefficients strongly depend
on the realization of the random impurity potential or the
shape of the island. Using random-matrix theory, we
calculate the thermopower and the thermoelectric kinetic
coefficient and study the statistics of their mesoscopic
fluctuations in the elastic cotunneling regime.
The fluctuations of the thermopower turn out to be
much larger than the average value.
\end{abstract}

 \maketitle

%%%%%%%%%%%%%%%%%%%%%%%%%%%%%%%%%%%%%%%%%%%%%%%%%%%%%%%%%%%%%%%%%%%%%%%%%%%%
%%%%%%%%%%%%%%%%%%%%%%%%%%%%%%%%%%%%%%%%%%%%%%%%%%%%%%%%%%%%%%%%%%%%%%%%%%%%

\section{Introduction}

Thermoelectric transport through various nanodevices has been the
subject of extensive experimental and theoretical studies for more
than two decades. The coherent propagation of electron waves in
clean nanostructured conductors leads to quantum size effects that
strongly affect the thermoelectric transport
coefficients;~\cite{VanHouten1992} the presence of electron-electron
interactions leads to additional renormalization
phenomena.~\cite{Kane1996,Fazio1998} In low-dimensional disordered
conductors, interference of diffusively scattered electron waves
weakens the screening of electron-electron interactions, leading to
anomalous, energy-dependent non-Fermi-liquid behavior of the
thermoelectric transport coefficients.~\cite{Catelani2005} All these
effects can be made visible explicitly using the tunability of
nanodevices, {\em e.g.}, by varying external gate potentials or
magnetic fields. Various practical applications based on
thermoelectric phenomena in nanostructures have been developed,
including thermometry, and
nanorefrigeration,~\cite{Giazotto2006,Muhonen2012,Courtois,Vasenko,Ozaeta}
and more generally thermoelectric
nanomachines.~\cite{DiSalvo1999,Shakouri2011,Whitney2014,Sothmann}

\begin{figure}[t]
\includegraphics[width=8cm]{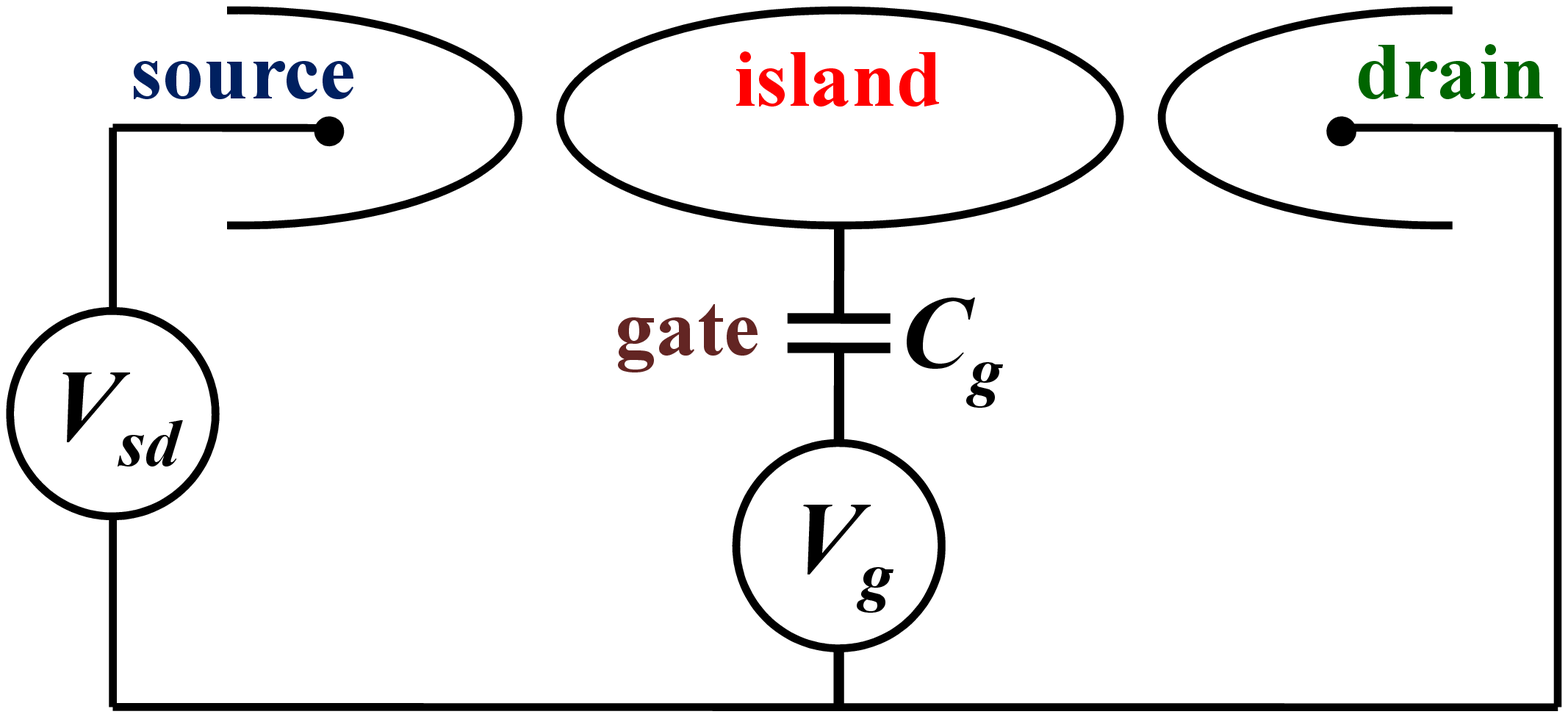}
\caption{(Color online) A sketch of a single-electron transistor.
The central island is connected to the source and drain electrodes
by tunnel junctions and capacitively coupled to the gate electrode
via the gate capacitance $C_g$. The total capacitance $C$ of the
island is given by the sum of the capacitances between the island
and each electrode.}\label{set}
\end{figure}

A prototypical device that manifests all the relevant aspects of
electron transport in nanostructures, {\em i.e.}, quantum size
effects, energy-dependent coherent propagation and electron-electron
interaction effects, is the single-electron transistor (SET). It
consists of an island (a quantum dot or a small metallic particle)
connected to two leads (source and drain) by small tunnel junctions.
The electrostatic potential on the island can be controlled
externally due a capacitive coupling between the island and a nearby
gate electrode with the capacitance $C_g$ (Fig.~\ref{set}). The
electrostatic energy cost of putting an extra electron on the island
is of the order of the so-called charging energy, $E_C \equiv e^2/
2C$, where $C$ is the total capacitance of the island. When the
temperature~$T$ and the applied source-drain voltage~$V_{sd}$ are
small, $eV_{sd}, T \ll E_C$, this charging effect results in the
so-called Coulomb blockade of the electron transport through the
island (see Ref.~\onlinecite{AleinerRev} for a review). Moreover, at
low temperatures the electronic phase coherence length is longer
than the typical dimensions of the island and as a result the
electronic motion is phase-coherent.

\begin{figure}[t]
\includegraphics[width=8cm]{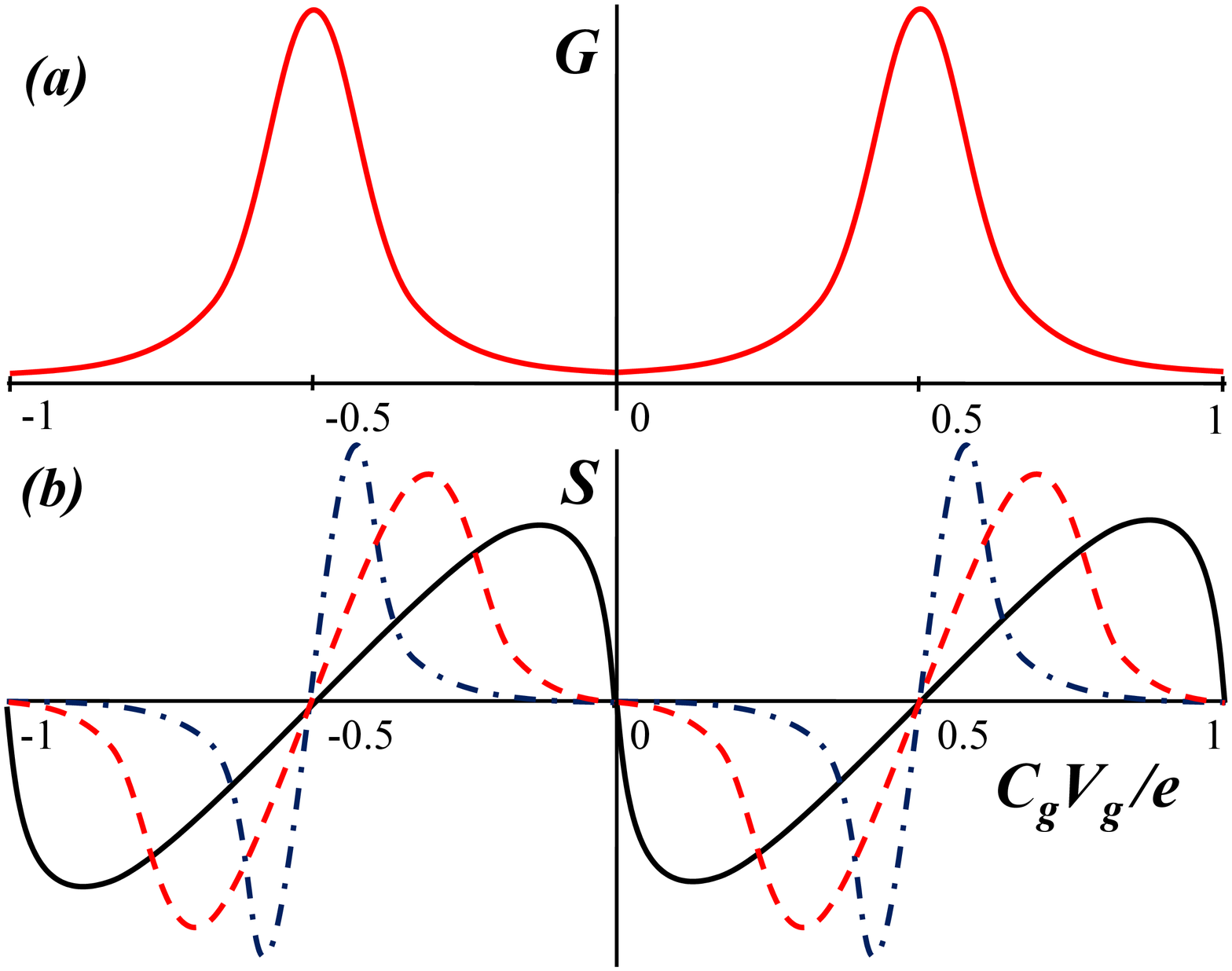}
\vspace{-3mm} \caption{(Color online) Schematic view of the
dependence of the SET's conductance~(a) and thermopower~(b) on the
dimensionless gate voltage $C_g V_g/e$. The three curves in
panel~(b) correspond to different temperatures, the black solid line
corresponding to the highest temperature, the blue dash-dotted line
to the lowest.}\label{TM} \vspace{-3mm}
\end{figure}

The number of electrons $N$ on the island that minimizes the
electrostatic energy $E_{el}(N)$, as well as the energy cost
$E_{el}(N\pm{1})-E_{el}(N)$ to add/remove an electron, depends on
the external gate voltage~$V_g$. For each~$N$, there is a particular
value of $V_g$, such that $E_{el}(N)=E_{el}(N+1)$, called the
degeneracy point. Then, starting from the state with $N$ electrons,
one electron can tunnel from the source to the island, and then
another electron tunnel from the island to the drain, restoring the
number of electrons on the island to~$N$. This so-called sequential
tunneling mechanism (when electrons tunnel one by one in and out of
the island, hence the term ``single-electron transistor'') leads to
a sequence of peaks in the dependence of the source-drain
linear-response conductance~$G$ on $V_g$, spaced at $e/C_g$,
schematically shown in Fig.~\ref{TM}a.

If $V_g$ is tuned away from the degeneracy point into the
Coulomb blockade valley,
the sequential-tunneling contribution to the conductance is
exponentially suppressed at low temperatures,%
\cite{Kulik1975,Beenakker} and a more important contribution to the
transport arises from the so-called cotunneling mechanism. It is due
to processes where an electron tunnels from the source to the
drain via a virtual intermediate state on the island. The energy of
this virtual state is higher than that of the initial and final
states by a large amount $\sim{E}_C$, so the tunneling amplitude is
small as $\sim{1}/E_C$. Yet, at low temperatures, this dominates over the
exponentially small sequential-tunneling contribution
$\sim\exp(-E_C/T)$.

If the internal state of the island (i.~e., the distribution of the
$N$~electrons over the single-particle energy levels on the island)
is different before and after the process, one speaks of inelastic
cotunneling, in the opposite case it is called elastic
cotunneling.\cite{Averin1989,Glazman1990,AN,ANbook} As the inelastic
cotunneling process involves creation of an electron-hole pair on
the island, the corresponding contribution to the linear-response
conductance vanishes at $T\to{0}$ ($\propto{T}^2$), while the
elastic one is temperature-independent. Thus, the latter dominates
the transport at very low temperatures.~\cite{footnote} An important
difference between inelastic and elastic cotunneling is that the
latter is sensitive to the coherent electron motion on the island
whereas the former is not. As a result, the elastic cotunneling
contribution to the SET's conductance shows strong mesoscopic
sample-to-sample fluctuations,~\cite{AG} the fluctuations being of
the same order as the average conductance. Moreover, the conductance
fluctuations of elastic cotunneling are so large that they dominate
the inelastic mechanism even at not too low temperatures, when the
average conductance value is already determined by inelastic
cotunneling.\cite{AG} The noise of the cotunneling current through
one or several tunnel-coupled quantum dots in the Coulomb blockade
regime was calculated in Ref.~\onlinecite{SL}.

The thermoelectric properties of SETs have been investigated in
part, both theoretically and experimentally. The thermoelectric
kinetic coefficient $G_T=I/\delta{T}$ is defined as the response of
the source-drain electric current $I$ to a small temperature
difference $\delta{T}$ between the source and the drain at zero
voltage, $V_{sd}=0$. The thermopower $S=-V_{sd}/\delta{T}$
determines the voltage response to~$\delta{T}$ at zero electric
current, $I=0$ (that is, with disconnected external circuit in
Fig.~\ref{set}). In the sequential tunneling regime, the thermopower
was predicted to exhibit periodic sawtooth oscillations as a
function of $V_g$ (see Ref.~\onlinecite{BS}), as shown schematically in
Fig.~\ref{TM}b (black solid line). This sawtooth behavior has been
observed experimentally, \cite{BSexp,Moller1998} however deviations
from it have also been seen.\cite{Dzurak1993,Dzurak1997} The latter
observations motivated the theoretical study of the inelastic
cotunneling contribution to the thermopower.\cite{TM} It was shown
that below some crossover temperature the thermopower in the valleys
of Coulomb blockade is supressed, the sawtooth behavior (black solid
line in Fig.~\ref{TM}b) is strongly modified at low temperatures
(blue dash-dotted line in Fig.~\ref{TM}b). Similar behavior was
later observed experimentally.\cite{TMexp} Taking into account the
inelastic cotunneling contribution also leads to the violation of
the Wiedemann-Franz law in a SET device, as was theoretically shown
in Ref.~\onlinecite{Kubala2008}.

The elastic cotunneling contribution to thermopower was discussed in
Refs.~\onlinecite{Koch2007,Billings2010}, but its statistics have
not been properly analyzed. At the same time, given the fact that
the elastic cotunneling regime gives rise to strong mesoscopic
conductance fluctuations, it appears crucial to study the statistics
of the thermoelectric coefficients in the elastic cotunneling
regime. The purpose of the present work is to perform such a study.
We consider thermoelectric transport through a small metallic island
containing many electrons, whose discrete single-particle energy
spectrum is characterized by the mean level spacing
$\Delta\ll{E}_C$. This spectrum, as well as the coherent electron
motion inside the dot, are assumed to be described by the orthogonal
ensemble of the random matrix theory, corresponding to the absence
of any external magnetic field. This assumption is valid as long as
$E_C$ is small compared to the Thouless energy of the island. At low
temperatures, $T\ll\sqrt{E_C \Delta}$, we can neglect the
contribution of the inelastic cotunneling.\cite{AN} Under these
assumptions, we determine (in the elastic cotunneling regime) the
full statistics of the thermoelectric kinetic coefficient~$G_T^{el}$
and of the thermopower, $S^{el}=G_T^{el}/G^{el}$, and show that the
fluctuations of $S^{el}$ are much larger than the average value.

The paper is organized as follows. In Section~\ref{results}, we
summarize the main results and discuss them qualitatively. We
specify the model in Section~\ref{model}. The detailed calculations
are presented in Section~\ref{calculations}. We summarize results in
Section~\ref{conclusions}.

%%%%%%%%%%%%%%%%%%%%%%%%%%%%%%%%%%%%%%%%%%%%%%%%%%%%%%%%%%%%%%%%%%%%%%%%%%%%
%%%%%%%%%%%%%%%%%%%%%%%%%%%%%%%%%%%%%%%%%%%%%%%%%%%%%%%%%%%%%%%%%%%%%%%%%%%%

\section{Qualitative discussion and summary of the main results}
\label{results}

\begin{figure}[t]
\includegraphics[width=8cm]{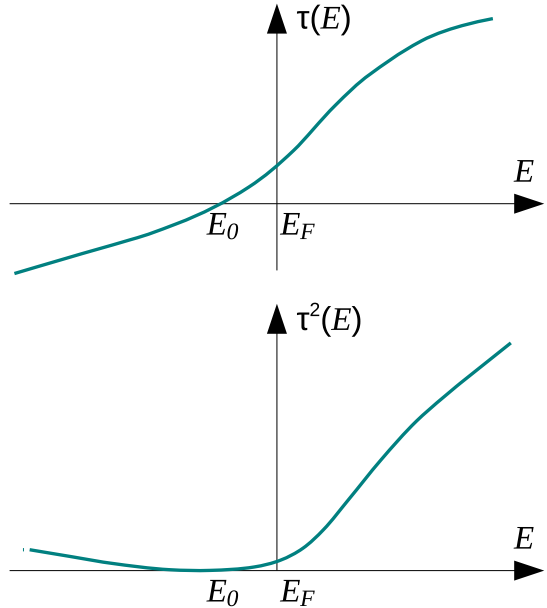}
\caption{(Color online) A sketch of possible behavior of the
function $\tau(E)$ (upper panel) and the corresponding $\tau^2(E)$
(lower panel), entering Eq.~(\ref{G=tau}).
The scale of the horizontal axis is $\sim{E}_C$, that of the
vertical axis is $\sim{1}$.}\label{fig:tau}
\end{figure}

Very generally, the linear response of charge and energy currents,
$I$~and~$J_E$, to the voltage and temperature differences,
$V$~and~$\delta{T}$, is determined by the $2\times{2}$ matrix of the
kinetic coefficients,
\begin{equation}\label{kinetic=}
\left(\begin{array}{c} I \\ J_E \end{array}\right)=
\left(\begin{array}{cc} G & G_T \\ K_V & K_T \end{array}\right)
\left(\begin{array}{c} V \\ \delta{T} \end{array}\right),
\end{equation}
where $K_V=-TG_T$ due to the Onsager symmetry holding under
time-reversal symmetry, thermopower $S=G_T/G$, and the thermal
conductance $K$ is given by $K=K_T-K_VG_T/G$. As will be shown in
Sec.~\ref{calculations}, for a given realization of the disorder or
of the shape of the island and a given value of the gate voltage,
the three independent kinetic coefficients in the regime of the
elastic cotunneling can be represented in the form
\begin{subequations}
\begin{align}
&G^{el}=\frac{G_sG_d}{4\pi{e}^2/\hbar}\,\frac\Delta{E_C}\,\tau^2(E_F),
\label{G=tau}\\
&G_T^{el}=-\frac{\pi^2T}{3e}\,\frac{G_sG_d}{4\pi{e}^2/\hbar}\,
\frac\Delta{E_C}\,\frac{d\tau^2(E_F)}{dE_F},\label{GT=tau}\\
&K_T^{el}=\frac{\pi^2T}{3e^2}\,\frac{G_sG_d}{4\pi{e}^2/\hbar}\,
\frac\Delta{E_C}\,\tau^2(E_F).
\end{align}
\end{subequations}
Here $G_s(G_d)$ is the conductance of the tunnel junction between
the island and the source (drain) electrode. $\tau(E_F)$ is a smooth
real dimensionless function of the Fermi energy~$E_F$ in the
electrodes, which depends on the microscopic realization of disorder
or the island shape, such as the one shown in Fig.~\ref{fig:tau}.
For a given
realization, it varies on a typical energy scale~$E_C$, and its
typical value is $\sim{1}$ (provided that the gate voltage is not too
close to a degeneracy point). Thus, for a given realization, the
kinetic coefficients satisfy the Mott formula for the thermopower
and the Wiedemann-Franz law for $K_T^{el}$. This is quite natural,
as these relations hold quite generally when the electron scattering
is elastic.\cite{Jonson1980}

As $G^{el}$ and $K_T^{el}$ depend on the realization via the same
quantity $\tau^2(E_F)$, their statistics is identical. It was found
in Ref.~\onlinecite{AG}, where the distribution function of the
elastic cotunneling electrical conductance $G^{el}$ was found
explicitly. Introducing the dimensionless variable $g=\tau^2(E_F)$,
see Eq.~(\ref{G=tau}), it coincides with the Porter-Thomas
distribution for the orthogonal ensemble,\cite{PT}
\begin{equation}
P(g)=\Theta(g)\sqrt{\frac{1-4x^2}{4\pi{g}}}\,
e^{-(1-4x^2)g/4},
\end{equation}
where $\Theta(g)$ is the Heaviside step function. Here, $x$ is the
rescaled gate voltage, such that $x=0$ corresponds to the center of
the Coulomb blockade valley, and $x=\pm{1}/2$ corresponds to the two
nearby degeneracy points. We restrict $x$ to the interval
$-1/2<x<1/2$, outside of which the dependence on the gate voltage
should be periodically repeated. The average elastic cotunneling
conductance value is given by Eq.(17) in Ref.~\onlinecite{AN},
\begin{equation} \label{Gelastic}
\langle G^{el} \rangle =
\frac{G_sG_d}{4\pi{e}^2/\hbar}\,\frac\Delta{E_C}\, \frac{2}{1 -
4x^2},
\end{equation}
and the fluctuations are indeed mesoscopically large, of the order
of the average conductance,
\begin{equation} \label{VarGelastic}
\sqrt{\langle (G^{el})^2 \rangle - \langle G^{el} \rangle^2} = \sqrt{2}\, \langle G^{el} \rangle.
\end{equation}

The quantity $G_T^{el}$ depends on the realization via
$d\tau^2(E_F)/dE_F$, therefore its statistics is different from that
of~$G^{el}$. However, these quantities are correlated. The
consequence of this fact for the statistics of the thermopower can
be understood from the following simple argument. The
function~$\tau(E)$ can have arbitrary sign, and it may even change
sign at some point $E=E_0$ (Fig.~\ref{fig:tau}).
In the vicinity of~$E_0$, it can be approximated as
$\tau(E)\approx{A}(E-E_0)$, where the coefficient $A$ is non-singular.
If $E_0$ happens to be close to~$E_F$, then
$G^{el}\propto\tau^2(E_F)\propto(E_F-E_0)^2$ is very small.
As $E_0$ is random and determined by the island, while $E_F$ is
determined by the electrodes, $E_0$~can be assumed uniformly
distributed in the vicinity of~$E_F$.
This immediately results in the $1/\sqrt{G^{el}}$ behavior of the
distribution function of $G^{el}\propto{A}^2(E_F-E_0)^2$ at
$G^{el}\to{0}$, as found in Ref.~\onlinecite{AG} in the orthogonal
ensemble. At the same time, the thermoelectric coefficient
$G_T^{el}\propto{d}\tau^2(E_F)/dE_F\propto(E_F-E_0)$, so the
thermopower $S^{el}=G_T^{el}/G^{el}\propto{1}/(E_F-E_0)$. For a
uniformly distributed $E_0$, this gives
$\alpha/(S^{el})^2$ for the asymptotics of the distribution function
at $S^{el}\to\pm\infty$ with some coefficient~$\alpha$. As the
coefficient is the same at $S^{el}\to+\infty$ and
$S^{el}\to-\infty$, such a distribution has a finite first moment
$\langle{S^{el}}\rangle$, but a divergent second moment
$\langle{(S^{el})}^2\rangle$, leading indeed to large mesoscopic
fluctuations of $S^{el}$. As we have just seen, these large
fluctuations are dominated by those realizations where $E_0$ and
$E_F$ happen to be close to each other, that is, the electrical
conductance is anomalously small.

These simple arguments are confirmed by the explicit calculation in
Sec.~\ref{calculations}, which gives the average elastic cotunneling
thermopower,
\begin{equation}\label{avS=}
\langle S^{el} \rangle = - \frac{\pi^2 T}{3 e E_C} \frac{4x}{1 -
4x^2},
\end{equation}
and divergent higher moments. Note that $\langle S^{el} \rangle=0$
at $x=0$, since in the valley center the system is
electron-hole-symmetric \emph{on the average}. However, for any
given realization, the electronic energy spectrum on the island does
not have any symmetry, so there is no reason for $S^{el}$ to vanish
at $x=0$ in any specific realization. Near the degeneracy points
$x=\pm 1/2$, Eq.~(\ref{avS=}) gives a divergence, but in this region
our theory does not work any more as the dominant contribution to
the transport comes from the sequential tunneling mechanism. The
full distribution function of the thermopower turns out to be a
simple Lorentzian, conveniently written in terms of the
dimensionless variable~$s$, such that $S^{el}=-[\pi^2T/(3eE_C)]s$:
\begin{equation}\label{Sdist=}
P(s)=\frac{\sqrt{4/3}}\pi\,\frac{1-4x^2}{4/3+[(1-4x^2)s-4x]^2}.
\end{equation}

\begin{figure}[t]
\includegraphics[width=8.5cm]{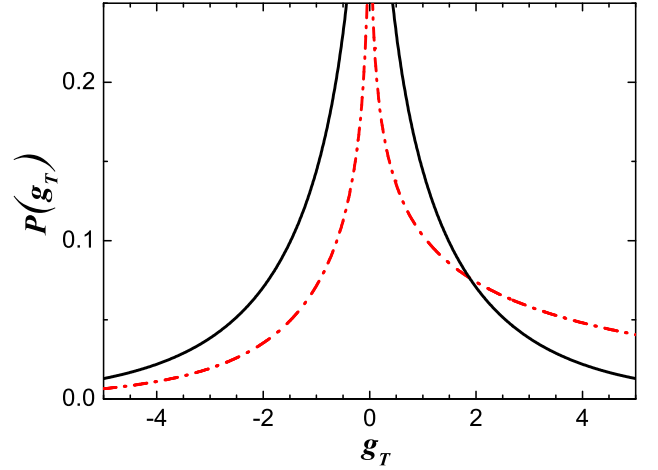}
\caption{(Color online) The distribution function $P(g_T)$ for $x=0$
(black solid line) and $x=0.3$ (red dash-dotted line).}
\label{distrib}
\end{figure}

We have also determined the full distribution function of $G_T^{el}$
written here in terms of a dimensionless variable
$g_T=E_C\,d\tau^2(E_F)/dE_F$ [see Eq.~(\ref{GT=tau})],
\begin{align}
P(g_T) = {}&\frac{\sqrt{3}}{4\pi}\, (1 - 4x^2)^2 \exp\left(
\frac{3}{2}\, x (1 - 4x^2)^2 g_T\right)\nonumber
\\
&\times K_0\!\left( \frac{\sqrt{3}}{4}\, (1 - 4x^2)^2
\sqrt{1 + 12 x^2}\, |g_T| \right),\label{distrib_f1}
\end{align}
where $K_0$ is the modified Bessel function.
The corresponding average value is given by
\begin{align}\label{G_Tav}
\langle G_T^{el} \rangle = -\frac{\pi^2T}{3e}\,
\frac{G_sG_d}{4\pi{e}^2/\hbar}\,\frac\Delta{E_C^2}\,
\frac{4x}{(1-4x^2)^2},
\end{align}
and higher moments are given by Eq.~(\ref{XYmoments=}).
Note that $\langle{G}_T^{el}/G^{el}\rangle=
2\langle{G}_T^{el}\rangle/\langle{G}^{el}\rangle$.
In Fig.~\ref{distrib} we plot $P(g_T)$ for two
different values of the dimensionless gate voltage $x=0,0.3$.
Note that changing the sign $x\to-x$ amounts to
$P(g_T)\to{P}(-g_T)$.

As $G^{el}$ and $G_T^{el}$ are correlated random variables, the full
information about their statistics at a given value of~$x$ is
provided by the joint distribution function, for which we have
obtained the following analytical expression:
\begin{align}
P(g,g_T) = {}&\frac{\sqrt{3}\,\Theta(g)}{8\pi{g}}\,(1 - 4x^2)^2
\exp\left(-\frac{1-4x^2}{4}\,g\right)\nonumber\\
{}&{}\times\exp
\left\{-\frac{3(1-4x^2)[4x(1-4x^2)g-g_T]^2}{16g}\right\}.\label{PggT=}
\end{align}
To characterize statistical correlations at different values of~$x$,
one should consider $g(x)$ and $g_T(x)$ as two correlated random
processes. They can be conveniently characterized in terms of
\begin{equation}\label{tauvep=}
\tau_x=\sqrt{g(x)}=\tau(E_F),\quad
\varepsilon_x=\frac{g_T(x)}{2\sqrt{g(x)}}
=E_C\,\frac{d\tau(E_F)}{dE_F},
\end{equation}
which turn out to be Gaussian random processes with zero
averages and pair correlators
\begin{subequations}
\begin{align}\label{avtautau=}
\langle\tau_{x_1}\tau_{x_2}\rangle={}&{}\frac{1}{2(x_1-x_2)}\,
\ln\left(\frac{1+2x_1}{1-2x_1}\,\frac{1-2x_2}{1+2x_2}\right),\\
\langle\tau_{x_1}\varepsilon_{x_2}\rangle={}&{}
\frac{1}{(x_1-x_2)(1-4x_2^2)}+{}\nonumber\\
&{}+\frac{1}{4(x_1-x_2)^2}
\ln\left(\frac{1+2x_1}{1-2x_1}\,\frac{1-2x_2}{1+2x_2}\right),\\
\langle\varepsilon_{x_1}\varepsilon_{x_2}\rangle={}&{}
\frac{1-2x_1^2-2x_2^2}{(x_1-x_2)^2(1-4x_1^2)(1-4x_2^2)}-{}\nonumber\\
&{}-\frac{1}{4(x_1-x_2)^3}
\ln\left(\frac{1+2x_1}{1-2x_1}\,\frac{1-2x_2}{1+2x_2}\right).
\label{avvepvep=}
\end{align}
\end{subequations}

The divergence of mesoscopic fluctuations of the thermopower,
found in the present work, originates from the fact that the
elastic cotunneling contribution to the electrical conductance,
calculated in the leading order in the tunneling couplings, has
too high a probability to vanish. Indeed, as discussed in the
paragraph preceding Eq.~(\ref{avS=}), both $G^{el}$ and $G_T^{el}$
may be small for some realizations, but it is easier for $G^{el}$
to have an anomalously small value, than for $G_T^{el}$, and then
their ratio $S^{el}=G_T^{el}/G^{el}$ becomes anomalously large.
One should recall, however, that the
leading-order elastic cotunneling is not the only contribution to
the conductance. There are other contributions (e.~g., the inelastic
cotunneling, or higher-order contributions to the elastic one),
which work as parallel conduction channels, so the conductance never
vanishes exactly. These contributions will cut off the divergence of
$\langle{S}^2\rangle$. Nevertheless, the fluctuations will
still be parametrically large. To estimate the magnitude of the
effect, let us assume that the elastic cotunneling conductance is
shunted by the inelastic one,\cite{AN}
$G^{in}\sim(G_sG_d\hbar/e^2)(T^2/E_C)^2$. The thermopower
fluctuations will be determined by those realizations which have
$G^{el}\sim{G}^{in}$, that is, $g\sim{T}^2/(E_c\Delta)\ll{1}$.
Then the typical value of $g_T\sim\sqrt{g}$, as seen from
Eq.~(\ref{PggT=}). Thus, we can estimate
$\sqrt{\langle{S}^2\rangle}/\langle{S}\rangle\sim{g}_T/g\sim
\sqrt{E_C\Delta}/T$. This factor is large precisely in the regime
when the elastic cotunneling dominates over the inelastic one.
Taking the values corresponding to the experiment of
Ref.~\onlinecite{TMexp}, $E_C=1.5$~meV, $\Delta=0.05$~meV,
$T\sim\Delta\approx{0}.6$~K, and $G_s = G_d = 0.012\,e^2/\hbar$,
we have $\sqrt{E_C\Delta}/T\sim{5}$. At lower temperatures,
$T\ll\Delta$, the inelastic cotunneling is suppressed even stronger,
so the fluctuations of the thermopower will be even larger.

\begin{figure}[t]
\includegraphics[width=8.5cm]{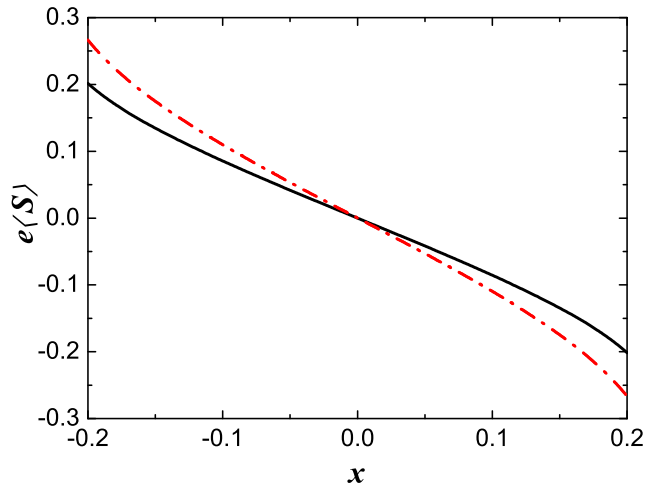}
\caption{(Color online) The averaged total thermopower with (black
solid line) and without (red dashed line) taking into account the
elastic cotunneling contiributions. The difference of two curves is
most visible in the interval $-0.2 < x < 0.2$. See the text for
details.} \label{FStot}
\end{figure}

At the same time, the effect of elastic cotunneling on the average
thermopower is not very dramatic. Even at $T\ll\sqrt{E_C\Delta}$
when the inelastic contributions $G^{in},G_T^{in}$ are small
compared to the typical values of the elastic ones,
$G^{in}\ll{G^{el}}$, $|G_T^{in}|\ll|G_T^{el}|$, the ratios
$S^{in}=G_T^{in}/G^{in}$ and
$\langle{S}^{el}\rangle=\langle{G}_T^{el}/G^{el}\rangle$ are of the
same order. Indeed, the average elastic cotunneling thermopower,
given by Eq.~(\ref{avS=}), differs from the inelastic one, given by
Eq.~(23) in Ref.~\onlinecite{TM},
\begin{equation}\label{SinTM}
S^{in} = - \frac{4 \pi^2 T}{5 e E_C} \frac{4x}{1 - 4x^2},
\end{equation}
just by a constant factor $12/5$.
To illustrate this effect, we include sequential tunneling and
inelastic cotunneling contributions to the conductance
($G^{sq},G^{in}$) and to the thermoelectric kinetic coefficient
($G_T^{sq},G_T^{in}$), and calculate the average of the total
thermopower,
\begin{equation}\label{avStot=}
\langle{S}\rangle=
\left\langle\frac{G_T^{sq}+G_T^{in}+G_T^{el}}{G^{sq}+G^{in}+G^{el}}\right\rangle,
\end{equation}
which is straightforwardly evaluated from the joint distribution
function~(\ref{PggT=}). Taking $G^{sq}$ and $G_T^{sq}$ from
Eqs.~(13) and (14) in Ref.~\onlinecite{TM}, respectively, and
$G^{in}$ and $G_T^{in}$ from Eq.~(10) in Ref.~\onlinecite{AN} and
Eq.~(22) in Ref.~\onlinecite{TM}, we plot in Fig.~\ref{FStot}
the average $\langle{S}\rangle$ for the values of the parameters
from the experiment of Ref.~\onlinecite{TMexp}, listed above, with
and without elastic cotunneling contributions. Thus, the qualitative
shape of the dependence of the average $\langle{S}\rangle$ on
the gate voltage is the same as in the elastic cotunneling case,
shown in Fig.~\ref{TM}b. However, for any specific realization of
the quantum dot, the dependence of $S$ will be different. In
particular, there is no reason why it would vanish exactly in the
center of the valley.

%

%%%%%%%%%%%%%%%%%%%%%%%%%%%%%%%%%%%%%%%%%%%%%%%%%%%%%%%%%%%%%%%%%%%%%%%%%%%%
%%%%%%%%%%%%%%%%%%%%%%%%%%%%%%%%%%%%%%%%%%%%%%%%%%%%%%%%%%%%%%%%%%%%%%%%%%%%

\section{The model}\label{model}

We model the single-electron transistor using the standard
Hamiltonian~\cite{AleinerRev},
\begin{subequations}
\begin{align}
&\hat{H}=\hat{H}_0+\hat{H}_{Ts}+\hat{H}_{Td},\\
&\hat{H}_0 = \sum_{\alpha=s,d}\sum_n
\xi_{\alpha,n} \hat{c}_{\alpha,n}^\dagger \hat{c}_{\alpha,n}
+ \sum_k \epsilon_k \hat{b}_k^\dagger \hat{b}_k
+ \hat{H}_C,\label{H0=}\\
&\hat{H}_{T\alpha} = \sum_{k,n} \left(
t_{\alpha,kn} \hat{c}^\dagger_{\alpha,n} \hat{b}_k
+ t_{\alpha,kn}^* \hat{b}^\dagger_k \hat{c}_{\alpha,n} \right).
\label{tunnel}
\end{align}
\end{subequations}
Here the subscript $\alpha=s,d$ labels the two electrodes
(source and drain, respectively), $n$ and $k$ label the
single-electron states in the leads and on the island,
respectively. For the sake of compactness, we suppress the
spin indices. As we will not consider spin-flip processes,
all subsequent calculations can be understood as performed
for a given spin projection, and the final expressions for
the transport coefficients will be multiplied by~2.
$\hat{c}_{\alpha,n}$ and $\hat{b}_k$ are the electron annihilation
operators for the corresponding states, and the corresponding
single-electron energies $\xi_{\alpha,n},\epsilon_k$ are measured
from the Fermi level.

$\hat{H}_C$ in Eq.~(\ref{H0=}) is the Coulomb interaction
Hamiltonian for electrons on the island, obtained from the standard
considerations.\cite{NBbook} Namely, the electrostatic energy
$E_{el}$ is assumed to be determined by the total electric charge
$Q=-Ne$ on the island, $E_{el}(Q)=\int^Q\varphi(Q')\,dQ'$, where
$\varphi(Q)=Q/C+C_gV_g/C$ is the electrostatic potential on the
island, and $C=C_g + C_s + C_d + C_i$ is the total capacitance of
the island, given by the sum of the capacitances to the gate
($C_g$), source ($C_s$), and drain ($C_d$) electrodes, as well as
the self-capacitance $C_i$ of the island. Thus, $\hat{H}_C$ can be
written as
\begin{equation}\label{HC=}
\hat{H}_C=E_C(\hat{N}^2-2\hat{N}C_g V_g/e),\quad
\hat{N}=\sum_k\left[\hat{b}_k^\dagger\hat{b}_k-\Theta(-\epsilon_k)\right].
\end{equation}
Here $\Theta(-\epsilon)$ is the Heaviside step function, so
$\hat{N}$ is the operator of the \textit{excess}
number of electrons on the island, the charge of the filled
Fermi sea at $\epsilon_k<0$ assumed to be compensated by the
neutralizing background. The degeneracy between states with
$N$ and $N+1$ on the island occurs when the dimensionless
gate voltage~$x$ is half-integer:
\begin{equation}
x\equiv\frac{C_g V_g}{e}=N+1/2.
\end{equation}
As the dependence of the transport coefficients on $V_g$ is
periodic, we can restrict our attention to the interval
$-1/2<x<1/2$, where the Coulomb energy is minimized by $N=0$. Thus,
$x$ measures the relative distance from the center of the Coulomb
blockade valley. It is convenient to introduce the Coulomb energy
cost of adding/removing one electron to/from the island,
\begin{equation}\label{Epm=}
E_\pm=E_{el}(N=\pm{1})-E_{el}(N=0)=E_C(1\mp{2x}).
\end{equation}

The matrix elements $t_{\alpha,kn}$ describe weak tunneling
coupling between the electrodes and the island. Analogously
to Ref.~\onlinecite{AG}, we assume that this coupling is due
to small overlap between the wave functions in the island and
in each electrode~$\alpha$, dominated by the vicinity of a
single point~$\mathbf{r}_\alpha$, where the island touches the
electrode~$\alpha$. Then the tunneling Hamiltonian can be
assumed to have the form
\begin{equation}
\hat{H}_{T\alpha} = t_\alpha\,
\hat\psi_\alpha^\dagger(\mathbf{r}_\alpha)
\hat\Psi(\mathbf{r}_\alpha)
+t_\alpha^*\,\hat\Psi^\dagger(\mathbf{r}_\alpha)\,
\hat\psi_\alpha(\mathbf{r}_\alpha),
\end{equation}
where $\hat\Psi(\mathbf{r})$ and $\hat\psi_\alpha(\mathbf{r})$
are the fermionic field operators for the electrons on the
island and in the contacts, respectively, and $t_\alpha$ are
the tunneling amplitudes incorporating all necessary
normalization factors.  Expanding the fermionic operators in
terms of the corresponding single-particle wave functions,
$\Psi_k(\mathbf{r})$ and $\psi_{\alpha,n}(\mathbf{r})$, as
\begin{equation}
\hat\Psi(\mathbf{r})=\sum_k\hat{b}_k\,\Psi_k(\mathbf{r}),
\quad
\hat\psi_\alpha(\mathbf{r})=\sum_n\hat{c}_{\alpha,n}
\psi_{\alpha,n}(\mathbf{r}),
\end{equation}
we obtain the following simple expression for the matrix
elements $t_{\alpha,kn}$:
\begin{equation}\label{takn=}
t_{\alpha,kn}=t_\alpha\,\Psi_k(\mathbf{r}_\alpha)\,
\psi_{\alpha,n}^*(\mathbf{r}_\alpha).
\end{equation}
The energies $\xi_{\alpha,n}$ are assumed to have continuous
spectra, so the leads are characterized by the local
densities of states (per spin)
\begin{equation}\label{nualpha=}
\nu_\alpha(\epsilon)=\sum_n|\psi_{\alpha,n}(\mathbf{r}_\alpha)|^2\,\delta(\epsilon-\xi_{\alpha,n}),
\end{equation}
assumed to be self-averaging and energy-independent. The energies
$\epsilon_k$ of the single-particle states on the island are
discrete with the mean level spacing~$\Delta$ ($2/\Delta$ being the
ensemble average of the single-electron density of states on the
island for both spin projections).
The island wave functions $\Psi_k(\mathbf{r}_\alpha)$ are
assumed to be real random variables, not correlated with the
energies~$\epsilon_k$, and corresponding to the elements of a random
orthogonal matrix uniformly distributed in the orthogonal group. To
the leading order in the matrix size, they can be treated as real
independent Gaussian random variables,\cite{AleinerRev} whose
statistics is entirely determined by the pair correlator
\begin{equation}\label{corrPsiPsi=}
%\langle\Psi^*_k(\mathbf{r}_\alpha)\,\Psi_{k'}(\mathbf{r}_{\alpha'})
%\rangle=
\langle\Psi_k(\mathbf{r}_\alpha)\,\Psi_{k'}(\mathbf{r}_{\alpha'})
\rangle=\delta_{kk'}\delta_{\alpha\alpha'},
\end{equation}
all normalization factors being absorbed in the tunneling
amplitudes $t_\alpha$ in Eq.~(\ref{takn=}).
Instead of $t_\alpha$, it is convenient to characterize each
tunneling contact by a physical quantity, such as its
average conductance (including the factor of 2 from spin),
\begin{equation}\label{Galpha=}
G_\alpha=2\,\frac{2\pi{e}^2}\hbar\,\frac{|t_\alpha|^2\nu_\alpha}\Delta.
\end{equation}

%%%%%%%%%%%%%%%%%%%%%%%%%%%%%%%%%%%%%%%%%%%%%%%%%%%%%%%%%%%%%%%%%%%%%%%%%%%%
%%%%%%%%%%%%%%%%%%%%%%%%%%%%%%%%%%%%%%%%%%%%%%%%%%%%%%%%%%%%%%%%%%%%%%%%%%%%

\section{Calculation}\label{calculations}

%%%%%%%%%%%%%%%%%%%%%%%%%%%%%%%%%%%%%%%%%%%%%%%%%%%%%%%%%%%%%%%%%%%%%%%%%%%%%%%%%%%%%%%%%%%%%%%%%
%%%%%%%%%%%%%%%%%%%%%%%%%%%%%%%%%%%%%%%%%%%%%%%%%%%%%%%%%%%%%%%%%%%%%%%%%%%%%%%%%%%%%%%%%%%%%%%%%

\subsection{Transport coefficients for a given realization}

Following Refs.~\onlinecite{Glazman1990,NBbook}, we start from
the Golden Rule expression for the source-drain current, which
represents the difference between the rate of electron transfer
from state $n$ on the source to the state $m$ on the drain
and the rate of the opposite process,
\begin{align}\label{current=FGR}
I=&-2e\sum_{n,m}\left[f_s(\xi_{s,n})-f_d(\xi_{d,m})\right]\nonumber\\
&{}\times
|M_{s,n\to{d},m}|^2\,\frac{2\pi}{\hbar}\,\delta(\xi_{s,n}-\xi_{d,m}),
\end{align}
where we used the fact that the cotunneling matrix element
$M_{d,m\to{s},n}=M_{s,n\to{d},m}^*$ and took into account the spin
degeneracy. In Eq.~(\ref{current=FGR}), $f_\alpha(\xi)$ is the
average occupation probability of the state with energy~$\xi$ on the
electrode~$\alpha$:
\begin{equation}
f_\alpha(\xi)=\frac{1}{1+e^{(\xi-\mu_\alpha)/T_\alpha}}.
\end{equation}
Focusing on the linear response to small chemical potential
and temperature differences, $\mu_s-\mu_d=-eV$, $T_s-T_d=\delta{T}$,
we can write
\begin{align}
&I=\frac{1}{\pi\hbar}\int(-e)\,\mathcal{T}(\xi)\left(-eV+\frac{\xi}{T}\,\delta{T}\right)
\left(-\frac{\partial{f}_{eq}}{\partial\xi}\right)d\xi,
\label{Ilinear=}\\
&\mathcal{T}(\xi)=4\pi^2\sum_{nm}
|M_{s,n\to{d},m}|^2\,\delta(\xi_{s,n}-\xi)\,\delta(\xi_{d,m}-\xi),
\label{Tdef=}
\end{align}
where $f_{eq}(\xi)=1/[1+\exp(\xi/T)]$. Using the same approach, one
can also find the energy current $J_E$ between the source and the
drain. The corresponding expression can be obtained from
Eq.~(\ref{Ilinear=}) by simply replacing the factor $(-e)$, which is
nothing but the charge transferred in a single tunneling event, by
the corresponding transferred energy $\xi$.

As will be seen below [Eq.~(\ref{T=})], $\mathcal{T}(\xi)$ is a
smooth function of $\xi$ varying on a typical scale of
$\xi\sim{E}_C$. At the same time, $-\partial{f}_{eq}/\partial\xi$ is
strongly peaked around zero on the scale $\xi\sim{T}$. Thus, at low
temperatures, the elastic cotunneling kinetic coefficients, appearing
in Eq.~(\ref{kinetic=}), can be approximated as
\begin{subequations}\begin{align}
&G^{el}=\frac{e^2}{\pi\hbar}\,\mathcal{T}(0),\label{GlowT=}\\
&G_T^{el}=-\frac{e}{\pi\hbar}\,\frac{\pi^2T}{3}\,\mathcal{T}'(0),
\label{GTlowT=}\\
&K_T^{el}=\frac{1}{\pi\hbar}\,\frac{\pi^2T}{3}\,\mathcal{T}(0),
\end{align}\end{subequations}
where we have used
\begin{equation}
\int\left(-\frac{\partial{f}_{eq}}{\partial\xi}\right)d\xi=1,\quad
\int\xi^2\left(-\frac{\partial{f}_{eq}}{\partial\xi}\right)d\xi
=\frac{\pi^2T^2}{3}.
\end{equation}
Note that in order to calculate $G_T^{el}$, one has to expand the
transmission function
$\mathcal{T}(\xi)\approx\mathcal{T}(0)+\xi\,\mathcal{T}'(0)$, as the
leading term vanishes due to parity $\xi\to-\xi$.

The cotunneling matrix element $M_{s,n\to{d},m}$ is evaluated
in the second-order perturbation theory in the tunneling
Hamiltonian $\hat{H}_T=\hat{H}_{Ts}+\hat{H}_{Td}$
as\cite{Glazman1990,NBbook}
\begin{equation}
M_{s,n\to{d},m}=\sum_v\frac{\langle\Phi|\hat{c}_{d,m}\hat{H}_T|v\rangle
\langle{v}|\hat{H}_T\hat{c}^\dagger_{s,n}|\Phi\rangle}{E_\Phi+\xi_{s,n}-E_v},
\end{equation}
where $\hat{c}^\dagger_{s,n}|\Phi\rangle$ is the initial state of
the system. It is conveniently represented as an extra electron on
top of some reference many-body state $|\Phi\rangle$, defined by the
occupation numbers of all single-particle states. The final state is
represented as $\hat{c}^\dagger_{d,m}|\Phi\rangle$, an extra
electron on top of {\em the same} reference state $|\Phi\rangle$,
which is the characteristic of the {\em elastic} cotunneling
process. The states $|v\rangle$~are virtual intermediate states with
energies~$E_v$. As $\hat{H}_T$ changes the number of electrons on
the island by one, the states $|v\rangle$ can belong to two sectors:
those with one more electron on the island (which is thus added to
some empty single-particle level~$k$), and those with one less
electron (which is thus removed from some filled single-particle
level~$k$).

Splitting the tunnel Hamiltonian Eq.~(\ref{tunnel}) as
\begin{subequations}\begin{align}
&\hat{H}_{T\alpha}=\hat{H}_{T\alpha-}+\hat{H}_{T\alpha+},\\
&\hat{H}_{T\alpha-}=
\sum_{k,n} t_{\alpha,kn} \hat{c}_{\alpha,n}^\dagger \hat{b}_k
=\hat{H}_{T\alpha+}^\dagger,
\end{align}\end{subequations}
we note that the first sector can be coupled to the initial
state only by the $\hat{H}_{Ts+}$ term, while the second
sector only by the $\hat{H}_{Td-}$ term. The energies of
the intermediate states in the two sectors are given by
\[
E_\Phi+E_++\epsilon_k,\quad
E_\Phi+E_--\epsilon_k+\xi_{s,n}+\xi_{d,m},
\]
respectively, where the Coulomb energies $E_\pm$ are defined
in Eq.~(\ref{Epm=}). Evaluation of the matrix elements gives
\begin{align}\label{M=}
M_{s,n\to{d},m}=&\sum_k t_{d,km}t_{s,kn}^*\times\nonumber\\
&{}\times\left(\frac{f_k}{E_--\epsilon_k+\xi_{d,m}}
-\frac{1-f_k}{E_++\epsilon_k-\xi_{s,n}}\right),
\end{align}
where $f_k$ is the occupation number of the single-electron
state~$k$ in the many-body state~$|\Phi\rangle$. It should be noted
that, strictly speaking, in a given reference state $|\Phi\rangle$,
$f_k$ is either 0 or~1. Then, to obtain the statistics of the
transport coefficients, averaging over different reference states
$|\Phi\rangle$ should also be performed, which results in the
probability of $f_k=1$ to be given by $f_{eq}(\epsilon_k)$. However,
this probability is different from 0 or 1 only in the range of
energies $|\epsilon_k|\sim{T}$, while, as will be seen later, the
sum over~$k$ in Eq.~(\ref{M=}) is contributed by a much wider range,
$|\epsilon_k|\sim{E}_C$. Thus, we approximate
$f_k=\Theta(-\epsilon_k)$. The error introduced by this
approximation is small by a factor $T/E_C\ll{1}$.

Substituting Eq.~(\ref{M=}) into Eq.~(\ref{Tdef=}) and using
Eqs.~(\ref{takn=}), (\ref{nualpha=}), and (\ref{Galpha=}), we obtain
\begin{equation}\label{T=}
\mathcal{T}(\xi)=\frac{\hbar^2G_sG_d}{4e^4}
\left|\sum_k\rho_k\left[
\frac{\Delta\,\Theta(\epsilon)}{E_++\epsilon-\xi}
-\frac{\Delta\,\Theta(-\epsilon)}{E_--\epsilon+\xi}\right]\right|^2,
\end{equation}
where we denoted
\begin{equation}
\rho_k=\Psi_k(\mathbf{r}_s)\,\Psi_k^*(\mathbf{r}_d).
\end{equation}

Using Eqs.~(\ref{GlowT=}) and (\ref{GTlowT=}), the transport
coefficients $G^{el}$ and $G_T^{el}$ can be expressed as
\begin{subequations}\begin{align}
&G^{el}=\frac{G_sG_d}{4\pi{e}^2/\hbar}\,X,\quad
G_T^{el}=-\frac{\pi^2T}{3e}\,\frac{G_sG_d}{4\pi{e}^2/\hbar}\,Y,\\
\label{Xdef=} &X=\sum_{k,k'}
\rho_k\rho_{k'}^*\,F(\epsilon_k)\,F(\epsilon_{k'}),\\
\label{Ydef=} &Y=\sum_{k,k'}\rho_k\rho_{k'}^*
\left[F(\epsilon_k)\,\tilde{F}(\epsilon_{k'})
+\tilde{F}(\epsilon_k)\,F(\epsilon_{k'})\right],\\
&F(\epsilon)=\frac{\Delta\,\Theta(\epsilon)}{E_++\epsilon}
-\frac{\Delta\,\Theta(-\epsilon)}{E_--\epsilon},\label{Fep=}\\
&\tilde{F}(\epsilon)
=\frac{\Delta\,\Theta(\epsilon)}{(E_++\epsilon)^2}
+\frac{\Delta\,\Theta(-\epsilon)}{(E_--\epsilon)^2}.\label{tFep=}
\end{align}\end{subequations}
The random variables $\tau_x,\varepsilon_x$, defined in
Eq.(\ref{tauvep=}), are represented as
\begin{equation}\label{tauvep=rho}
\tau_x=\sqrt{\frac{E_C}\Delta}\sum_k\rho_k\,F(\epsilon_k),
\quad
\varepsilon_x=\sqrt{\frac{E_C^3}\Delta}
\sum_k\rho_k\,\tilde{F}(\epsilon_k).
\end{equation}

%%%%%%%%%%%%%%%%%%%%%%%%%%%%%%%%%%%%%%%%%%%%%%%%%%%%%%%%%%%%%%%%%%%%%%%%%%%%%%%%%%%%%%%%%%%%%%%%%
%%%%%%%%%%%%%%%%%%%%%%%%%%%%%%%%%%%%%%%%%%%%%%%%%%%%%%%%%%%%%%%%%%%%%%%%%%%%%%%%%%%%%%%%%%%%%%%%%

\subsection{Averaging over realizations}

The statistics of the transport coefficients will be obtained from
the joint moments of $G^{el}$ and $G_T^{el}$, as it was done in
Ref.~\onlinecite{AG} for~$G^{el}$ alone. Thus, we need to average
products of $X$'s and $Y$'s [defined in Eqs.~(\ref{Xdef=}),
(\ref{Ydef=})] over the energies $\epsilon_k$ and wave function
amplitudes $\Psi_k(\mathbf{r}_\alpha)$. This task is facilitated by
the following two considerations.

(i)~Strictly speaking, the positions of the random energy levels
$\epsilon_k$ are correlated.\cite{MehtaBook} However, these
correlations occur on the energy scale~$\Delta$. Thus, for any
smooth function of energy, $\mathcal{F}(\epsilon)$, varying on the
scale $\epsilon\sim{E}_C$, the sum over $\epsilon_k$ will be
replaced by the integration over~$\epsilon$,
\begin{equation}\label{sumk=}
\sum_k\mathcal{F}(\epsilon_k)\to\int\mathcal{F}(\epsilon)\,
\frac{d\epsilon}\Delta.
\end{equation}
This approximation introduces an error which is small by a factor
$\Delta/E_C$. Of particular importance for the future calculations
will be the following three integrals:
\begin{subequations}
\begin{align}
\mathcal{J}_1 &= \int F^2(\epsilon)\,\frac{d\epsilon}\Delta
= \frac{\Delta}{E_+} + \frac{\Delta}{E_-}, \label{J1=}\\
\mathcal{J}_2 &= \int F(\epsilon)\, \tilde{F}(\epsilon)\,
\frac{d\epsilon}\Delta
 = \frac{\Delta}{2E_+^2} - \frac{\Delta}{2E_-^2},\label{J2=}\\
\mathcal{J}_3 &= \int \tilde{F}^2(\epsilon)\,\frac{d\epsilon}\Delta
= \frac{\Delta}{3E_+^3} + \frac{\Delta}{3E_-^3}.\label{J3=}
\end{align}
\end{subequations}

(ii)~When averaging over the wave function amplitudes
$\Psi_k(\mathbf{r}_\alpha)$ (which is independent of the averaging
over~$\epsilon_k$) using Eq.~(\ref{corrPsiPsi=}) in the orthogonal
ensemble, strictly speaking, all possible pairings should be taken,
in accordance with Wick's theorem for Gaussian random variables.
However, the amplitudes enter $X$ and $Y$ via a combination
$\rho_k=\Psi_k(\mathbf{r}_s)\,\Psi_k^*(\mathbf{r}_d)$. Then, for a
product $\rho_{k_1}\ldots\rho_{k_{2n}}$ [a product of an odd number of
factors always vanishes because $\Psi_k(\mathbf{r}_s)$ and
$\Psi_{k'}(\mathbf{r}_d)$ are uncorrelated], those pairings are more
important, where
$\Psi_{k_1}(\mathbf{r}_s)\ldots\Psi_{k_{2n}}(\mathbf{r}_s)$ are
paired exactly in the same way as
$\Psi_{k_1}(\mathbf{r}_d)\ldots\Psi_{k_{2n}}(\mathbf{r}_d)$, as it
gives the minimal number of constraints on the indices. This happens
because each summation is transformed into integration over a range
of $\epsilon\sim{E}_C$, and thus produces a large factor
$\sim{E}_C/\Delta$, as discussed in the previous paragraph.

To illustrate this fact, consider the average
\begin{align}
\langle{Y}^2\rangle={}&4\sum_{k_1\ldots{k}_4} \langle
\rho_{k_1}\rho_{k_2}\rho_{k_3}\rho_{k_4}\rangle\,
F(\epsilon_{k_1})\,F(\epsilon_{k_2})\,
\tilde{F}(\epsilon_{k_3})\,\tilde{F}(\epsilon_{k_4})\nonumber\\
%=\left(\delta_{k_1k_2}\delta_{k_3k_4}
%+\delta_{k_1k_3}\delta_{k_2k_4}\right)
%\left(\delta_{k_1k_2}\delta_{k_3k_4}
%+\delta_{k_1k_3}\delta_{k_2k_4}\right)
={}&4\sum_{k_1\ldots{k}_4}
F(\epsilon_{k_1})\,F(\epsilon_{k_2})\,
\tilde{F}(\epsilon_{k_3})\,\tilde{F}(\epsilon_{k_4})
\times{}\nonumber\\
&\times
\left[\delta_{k_1k_2}\delta_{k_3k_4}+\delta_{k_1k_3}\delta_{k_2k_4}
+2\delta_{k_1k_2}\delta_{k_1k_3}\delta_{k_1k_4}\right]\nonumber\\
={}&4\sum_{k,k'}F^2(\epsilon_k)\,\tilde{F}^2(\epsilon_{k'})
+{}\nonumber\\
&+4\sum_{k,k'}F(\epsilon_k)\,\tilde{F}(\epsilon_k)\,
F(\epsilon_{k'})\,\tilde{F}(\epsilon_{k'})+{}\nonumber\\
&+8\sum_kF^2(\epsilon_k)\,\tilde{F}^2(\epsilon_k)\nonumber\\
={}&4(\mathcal{J}_1\mathcal{J}_3+\mathcal{J}_2^2)[1+O(\Delta/E_C)].
\end{align}
The third term contains a single sum instead of a double sum, so
its contribution is smaller by a factor $\Delta/E_C$. Neglecting
those pairings, which produce extra constraints on the indices, is
equivalent to treating $\rho_k$'s as real independent Gaussian
random variables with the pair correlator
\begin{equation}
\langle\rho_k\rho_{k'}\rangle=\delta_{kk'}.
\end{equation}

Armed with this knowledge,
we are now ready to calculate an arbitrary joint moment,
\begin{align}
\langle{X}^pY^q\rangle={}&2^q\sum_{k_1\ldots{k}_{2p+2q}}
\langle\rho_{k_1}\ldots\rho_{k_{2p+2q}}\rangle
\nonumber\\ &{}\times
%F(\epsilon_{k_1})\ldots{F}(\epsilon_{k_{2p+q}})\,
%\tilde{F}(\epsilon_{k_{2p+q+1}})\ldots
%\tilde{F}(\epsilon_{k_{2p+2q}})
F_1\ldots{F}_{2p+q}\tilde{F}_{2p+q+1}\ldots
\tilde{F}_{2p+2q},
\end{align}
where we denoted $F(\epsilon_{k_i})=F_i$,
$\tilde{F}(\epsilon_{k_i})=\tilde{F}_i$ for compactness.
Evaluation of the average
$\langle\rho_{k_1}\ldots\rho_{k_{2p+2q}}\rangle$
amounts to summation over all possible pairings of $\rho_k$'s.
Each pairing of $\rho_k$'s induces a pairing of $F_k$'s and
$\tilde{F}_k$'s, then the subsequent summation over
the corresponding $k$ index is performed independently from
other indices, thereby producing a factor $\mathcal{J}_1$
for $FF$, $\mathcal{J}_2$ for $F\tilde{F}$, and $\mathcal{J}_3$
for $\tilde{F}\tilde{F}$, according to Eqs.~(\ref{sumk=}) and
(\ref{J1=})--(\ref{J3=}).

\begin{figure}
\includegraphics[width=8cm]{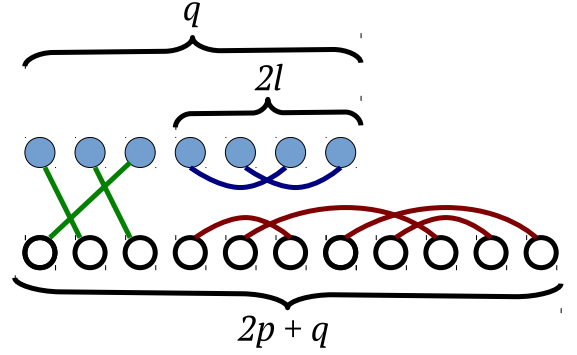}
\caption{(Color online) An example of pairing
$2p+q$ $F$'s (shown by empty circles) and $q$ $\tilde{F}$'s (filled
circles). }\label{fig:pairing}
\end{figure}

Let us classify all possible pairings by the number $2l$ of all
$\tilde{F}$'s which are paired among themselves, so that the
remaining $q-2l$ $\tilde{F}$'s are paired with $F$'s, and the
remaining $2p+2l$ $F$'s are also paired among themselves, as
illustrated graphically in Fig.~\ref{fig:pairing}. Obviously,
$0\leqslant{l}\leqslant{q}/2$, and all pairings with a given~$l$
give a factor
$\mathcal{J}_1^{p+l}\mathcal{J}_2^{q-2l}\mathcal{J}_3^l$. To
determine the combinatorial coefficient, we first note that there
are $q!/[(2l)!\,(q-2l)!]$ ways to choose $2l$ $\tilde{F}$'s out
of~$q$, and $(2p+q)!/[(q-2l)!\,(2p+2l)!]$ ways to choose $2p+2l$
$F$'s out of $2p+q$. Then, there are $(q-2l)!$ ways to pair up
$q-2l$ $F$'s with $q-2l$ $\tilde{F}$'s, $(2l-1)!!=(2l)!/(2^ll!)$
ways to pair up the $2l$ $\tilde{F}$'s among themselves, and
$(2p+2l-1)!!=(2p+2l)!/[2^{p+l}(p+l)!]$ ways to pair up the $2p+2l$
$F$'s. As a result,
\begin{equation}\label{XYmoments=}
\langle{X}^pY^q\rangle=\sum_{0\leqslant{l}\leqslant{q}/2}
\frac{2^{q-2l-p}\,q!\,(2p+q)!}{l!\,(q-2l)!\,(p+l)!}\,
\mathcal{J}_1^{p+l}\mathcal{J}_2^{q-2l}\mathcal{J}_3^l.
\end{equation}
From the joint moments, the full statistics can be reconstructed
following the standard procedure (given in detail in the Appendix).

%%%%%%%%%%%%%%%%%%%%%%%%%%%%%%%%%%%%%%%%%%%%%%%%%%%%%%%%%%%%%%%%%%%%%%%%%%%%%%%%%%%%%%%%%%%%%%%%%
%%%%%%%%%%%%%%%%%%%%%%%%%%%%%%%%%%%%%%%%%%%%%%%%%%%%%%%%%%%%%%%%%%%%%%%%%%%%%%%%%%%%%%%%%%%%%%%%%

\section{Conclusions}\label{conclusions}

We have developed a quantitative theory of thermoelectric transport
in a single electron transistor, consisting of a quantum dot weakly coupled
to two electronic leads and capacitively coupled to a gate electrode, in
the regime of elastic cotunneling. In this regime, the transport coefficients
strongly depend on the realization of the random impurity potential or the
shape of the island. We assumed the quantum dot wave functions to be Gaussian
random variables and used the random-matrix theory for the orthogonal
ensemble (i.e. in the absence of the magnetic field).

The distribution function of the conductance $G$ was previously obtained by
Aleiner and Glazman in Ref.~\onlinecite{AG}. We have extended this result and calculated the distributions of the thermopower $S$, the
thermoelectric
kinetic coefficient $G_T$, and the joint distribution function of the conductance and
thermoelectric kinetic coefficient as functions of the gate voltage. Statistical
correlations of $G$ and $G_T$ at different values of the gate potential were
also calculated.

Finally, we have calculated the average elastic cotunneling values of the
thermopower and thermoelectric kinetic coefficient and the average values
of all moments of $G_T$. We have shown that the second and higher moments
of the thermopower diverge, which leads to large mesoscopic fluctuations of the
elastic cotunneling thermopower. This divergence is cut off by taking into account
the inelastic cotunneling contribution, or higher-order contributions to the elastic one.
Nevertheless, the fluctuations will still be parametrically large.

We have estimated the magnitude of these fluctuations, taking into account the experimental
parameters from Ref.~\onlinecite{TMexp}, $\sqrt{\langle{S}^2\rangle}/\langle{S}\rangle
\sim 5$. Therefore for any specific realisation of the quantum dot, the dependence
of $S$ on the gate voltage will be different from $\langle{S}\rangle$. In particular,
there is no reason why it would vanish exactly in the center of the Coulomb blockade valley.

\section{Acknowledgements}

The authors are grateful to D.V. Averin, V. Bubanja, R. Whitney, and
C. Winkelmann for helpful discussions. This work was supported by
European Union Seventh Framework Programme (FP7/2007--2013) under
grant agreement ``INFERNOS'' No.~308850, as well as by Institut
Universitaire de France.

%%%%%%%%%%%%%%%%%%%%%%%%%%%%%%%%%%%%%%%%%%%%%%%%%%%%%%%%%%%%%%%%%%%%%%%%%%%%%%%%%%%%%%%%%%%%%%%%%
%%%%%%%%%%%%%%%%%%%%%%%%%%%%%%%%%%%%%%%%%%%%%%%%%%%%%%%%%%%%%%%%%%%%%%%%%%%%%%%%%%%%%%%%%%%%%%%%%

\appendix

\section{Statistics from joint moments}
\renewcommand{\theequation}{A\arabic{equation}}
\setcounter{equation}{0}

From the joint moments~(\ref{XYmoments=}), we first
reconstruct the characteristic function:
\begin{align}
\chi(u,v)={}&\left\langle{e}^{-iuX-ivY}\right\rangle
={}\sum_{p,q=0}^\infty\frac{(-iu)^p(-iv)^q}{p!\,q!}\,
\langle{X}^pY^q\rangle\nonumber\\
={}&\sum_{p=0}^\infty\sum_{l=0}^\infty\sum_{q=2l}^\infty
\frac{(2p+q)!}{p!\,(p+l)!\,l!\,(q-2l)!}\,\nonumber\\
&{}\times(-iu\mathcal{J}_1/2)^p
(-v^2\mathcal{J}_1\mathcal{J}_3)^l(-2iv\mathcal{J}_2)^{q-2l}
\nonumber\\
={}&\frac{1}{1+2iv\mathcal{J}_2}
\sum_{p=0}^\infty\sum_{l=0}^\infty
\frac{(2p+2l)!}{p!\,(p+l)!\,l!}\,\nonumber\\
&{}\times\left[\frac{-iu\mathcal{J}_1/2}%
{(1+2iv\mathcal{J}_2)^2}\right]^p
\left[\frac{-v^2\mathcal{J}_1\mathcal{J}_3}%
{(1+2iv\mathcal{J}_2)^2}\right]^l\nonumber\\
={}&\frac{1}{1+2iv\mathcal{J}_2}\sum_{s=0}^\infty
\frac{(2s)!}{(s!)^2}\left[
\frac{-iu\mathcal{J}_1/2-v^2\mathcal{J}_1\mathcal{J}_3}%
{(1+2iv\mathcal{J}_2)^2}\right]^s\nonumber\\
={}&\frac{1}{\sqrt{(1+2iv\mathcal{J}_2)^2+2iu\mathcal{J}_1
+4v^2\mathcal{J}_1\mathcal{J}_3}}.
\end{align}
The sums were calculated using the following relations:
\begin{subequations}\begin{align}
&\sum_{n=0}^\infty\frac{(m+n)!}{n!}\,z^n=\frac{m!}{(1-z)^{m+1}},\\
&\sum_{m,n=0}\frac{\mathcal{F}(m+n)}{m!\,n!}\,x^my^n=
\sum_{N=0}^\infty\sum_{k=0}^N\frac{\mathcal{F}(N)}{k!\,(N-k)!}\,
x^ky^{N-k}\nonumber\\
&{}\hspace*{3.5cm}=
\sum_{N=0}^\infty\frac{\mathcal{F}(N)}{N!}\,(x+y)^N,\\
&\sum_{n=0}^\infty\frac{(2n)!}{(n!)^2}\,z^n=
\frac{1}{\sqrt{1-4z}}.
\end{align}\end{subequations}
From the characteristic function, the probability distributions
can be determined. For the conductance,
\begin{equation}
P(X)=\int\frac{du}{2\pi}\,e^{iuX}\,\chi(u,0)
=\frac{\Theta(X)\,e^{-X/(2\mathcal{J}_1)}}%
{\sqrt{2\pi\mathcal{J}_1X}}
\end{equation}
coincides with the result of Ref.~\onlinecite{AG} for the
orthogonal ensemble. For the thermoelectric kinetic coefficient,
\begin{align}
P(Y)={}&\int\frac{dv}{2\pi}\,e^{ivY}\,\chi(0,v)\nonumber\\
={}&\frac{1}{2\pi\sqrt{\mathcal{J}_1\mathcal{J}_3-\mathcal{J}_2^2}}\,
\exp\left(\frac{\mathcal{J}_2}%
{\mathcal{J}_1\mathcal{J}_3-\mathcal{J}_2^2}\,\frac{Y}2\right)
\nonumber\\
&{}\times K_0\!\left(\frac{\sqrt{\mathcal{J}_1\mathcal{J}_3}}%
{\mathcal{J}_1\mathcal{J}_3-\mathcal{J}_2^2}\,\frac{Y}2\right),
\label{Ydist=}
\end{align}
where $K_0$ is the modified Bessel function.
Here it was important that
$\lambda=\mathcal{J}_1\mathcal{J}_3/\mathcal{J}_2^2
=1+1/(12x^2)\geqslant{1}$, so the following relation could be used:
\begin{align}
&\int\limits_{-\infty}^{\infty} \frac{d q}{2\pi}\,
\frac{e^{i q z}}{\sqrt{ (1 + iq)^2 - \lambda (iq)^2 }}\nonumber
\\
&= \frac{1}{\pi \sqrt{\lambda - 1}}
\exp\left( \frac{z}{ \lambda - 1} \right)
K_0\!\left( |z| \frac{\sqrt{\lambda}}{\lambda - 1} \right).
\end{align}
In combination with Eqs.~(\ref{J1=})--(\ref{J3=}) and with the
facts that
$\mathcal{J}_1\mathcal{J}_3=(\Delta^2/E_C^4)(4/3)(1+12x^2)/(1-4x^2)^4$,
and
$\mathcal{J}_1\mathcal{J}_3-\mathcal{J}_2^2=(\Delta^2/E_C^4)(4/3)/(1-4x^2)^4$,
Eq.~(\ref{Ydist=}) gives Eq.~(\ref{distrib_f1}).
The joint probability distribution is given by
\begin{align}
P(X,Y)={}&\int\frac{du}{2\pi}\,\frac{dv}{2\pi}\,e^{iuX+ivY}\,
\chi(u,v)=\nonumber\\
={}&\frac{\Theta(X)}{\sqrt{2\pi\mathcal{J}_1X}}
\int\frac{dv}{2\pi}\times{}\nonumber\\
&{}\times{e}^{ivY-[(1+2iv\mathcal{J}_2)^2
+4v^2\mathcal{J}_1\mathcal{J}_3]X/(2\mathcal{J}_1)}\nonumber\\
={}&\frac{\Theta(X)}%
{4\pi{X}\sqrt{\mathcal{J}_1\mathcal{J}_3-\mathcal{J}_2^2}}\times{}\nonumber\\
&{}\times\exp\left[-\frac{X}{2\mathcal{J}_1}
-\frac{(2\mathcal{J}_2X-\mathcal{J}_1Y)^2}%
{8(\mathcal{J}_1\mathcal{J}_3-\mathcal{J}_2^2)\mathcal{J}_1X}\right].
\end{align}
From this, the distribution function for the thermopower can be
obtained by introducing the variable $Z=Y/X$,
\begin{align}
P(Z)={}&{}\int\delta(Z-Y/X)\,P(X,Y)\,dX\,dY\nonumber\\
={}&{}\frac{(2\mathcal{J}_1/\pi)
\sqrt{\mathcal{J}_1\mathcal{J}_3-\mathcal{J}_2^2}}%
{4(\mathcal{J}_1\mathcal{J}_3-\mathcal{J}_2^2)
+(2\mathcal{J}_2-\mathcal{J}_1Z)^2},
\end{align}
which gives Eq.~(\ref{Sdist=}).

Finally, to describe the correlations of the random processes
$\tau_x,\varepsilon_x$ at different~$x$, it is sufficient to use
representation~(\ref{tauvep=rho}), and calculate the
characteristic functional
\begin{subequations}
\begin{align}
\mathcal{X}[u(x),v(x)]={}&{}
\left\langle\exp\left\{i\int[u(x)\,\tau_x+v(x)\,\varepsilon_x]\,
dx\right\}\right\rangle=\nonumber\\
={}&{}\exp\left\{-\frac{1}{2}\int{d}x_1\,dx_2\,
\mathcal{K}(x_1,x_2)\right\},\label{calX=}
\end{align}
\begin{align}
\mathcal{K}(x_1,x_2)={}&{}
u(x_1)\,u(x_2)\,\frac{E_C}\Delta\sum_k
F_{x_1}(\epsilon_k)\,F_{x_2}(\epsilon_k)+{}\nonumber\\
&{}+2u(x_1)\,v(x_2)\,\frac{E_C^2}\Delta\sum_k
F_{x_1}(\epsilon_k)\,\tilde{F}_{x_2}(\epsilon_k)+{}\nonumber\\
&{}+v(x_1)\,v(x_2)\,\frac{E_C^3}\Delta\sum_k
\tilde{F}_{x_1}(\epsilon_k)\,\tilde{F}_{x_2}(\epsilon_k),
\label{calK=}
\end{align}
\end{subequations}
where the subscripts at $F(\epsilon_k)$, $\tilde{F}(\epsilon_k)$
indicate that the values $E_+,E_-$, entering in Eqs.~(\ref{Fep=}),
Eqs.~(\ref{tFep=}), should be taken at the corresponding value
of~$x$, see Eq.~(\ref{Epm=}). Eqs.~(\ref{calX=}), (\ref{calK=})
represent the characteristic functional of a pair of Gaussian random
processes, whose pair correlators are given by
Eqs.~(\ref{avtautau=})--(\ref{avvepvep=}), obtained by evaluation of
the sums in Eq.~(\ref{calK=}) using the rule~(\ref{sumk=}).

%%%%%%%%%%%%%%%%%%%%%%%%%%%%%%%%%%%%%%%%%%%%%%%%%%%%%%%%%%%%%%%%%%%%%%%%%%%%%%%%%%%%%%%%%%%%%%%%%
%%%%%%%%%%%%%%%%%%%%%%%%%%%%%%%%%%%%%%%%%%%%%%%%%%%%%%%%%%%%%%%%%%%%%%%%%%%%%%%%%%%%%%%%%%%%%%%%%

\end{document}